\documentclass[preprint,amsmath,eqsecnum,aps,floats,groupedaddress]{revtex4}

\usepackage{graphicx}% Include figure files
\usepackage{dcolumn}% Align table columns on decimal point
\usepackage{bm}% bold math
\usepackage[closeFloats]{fltpage}
\usepackage{nopageno}
\usepackage{color}



\begin{document}

\title{Creation and control of a two-dimensional electron liquid at the bare SrTiO$_3$ surface}

\author{W. Meevasana$^{1, 2, 3, 4, 5*}$, P.D.C. King$^{3*}$, R.H. He$^{1,2}$, S.-K. Mo$^{1, 6}$, M. Hashimoto$^{1, 6}$, A. Tamai$^{3}$, P. Songsiriritthigul$^{4, 5}$, F. Baumberger$^{3}$, Z.-X. Shen$^{1,2\S}$}

\affiliation {$^1$Departments of Physics and Applied Physics,
Stanford University, CA 94305, USA}

\affiliation {$^2$Stanford Institute for Materials and Energy
Sciences, SLAC National Accelerator Laboratory, 2575 Sand Hill
Road, Menlo Park, CA 94025, USA}

\affiliation {$^3$School of Physics and Astronomy, University of
St Andrews, North Haugh, St. Andrews, Fife KY16 9SS, UK}

\affiliation {$^4$School of Physics, Suranaree University of
Technology and Synchrotron Light Research Institute, Nakhon
Ratchasima, 30000, Thailand}

\affiliation {$^5$Thailand Center of Excellence in Physics, CHE,
Bangkok, 10400, Thailand }

\affiliation {$^6$Advanced Light Source, Lawrence Berkeley
National Lab, Berkeley, CA 94720, USA}

\affiliation {$^*$These authors contributed equally to this work.}
\affiliation {$^\S$To whom correspondence should be addressed;
E-mail: zxshen@stanford.edu}

\date{\today}

\maketitle

%\begin{raggedright}
\parindent=0.5in

%\begin{abstract}
\textbf{Many-body interactions in transition-metal oxides give
rise to a wide range of functional properties, such as
high-temperature superconductivity \cite{HighTc:Bednorz}, colossal
magnetoresistance \cite{LMO:Helmolt}, or multiferroicity
\cite{multiferroic:Kimura}. The seminal recent discovery of a
two-dimensional electron gas (2DEG) at the interface of the
insulating oxides LaAlO$_3$ and SrTiO$_3$ \cite{LAO:Hwang}
represents an important milestone towards exploiting such
properties in all-oxide devices \cite{Oxide:Takagi}. This
conducting interface shows a number of appealing properties,
including a high electron mobility \cite{LAO:Hwang, Thiel},
superconductivity \cite{scinterface:Reyren}, and large
magnetoresistance \cite{magnetic:Brinkman} and can be patterned on
the few-nanometer length scale. However, the microscopic origin of
the interface 2DEG is poorly understood. Here, we show that a
similar 2DEG, with an electron density as large as
$8\times10^{13}$~cm$^{-2}$, can be formed at the bare SrTiO$_3$
surface. Furthermore, we find that the 2DEG density can be
controlled through exposure of the surface to intense ultraviolet
(UV) light. Subsequent angle-resolved photoemission spectroscopy
(ARPES) measurements reveal an unusual coexistence of a light
quasiparticle mass and signatures of strong many-body
interactions.}

%\end{abstract}
%\pacs{}

It has been known for decades that strong electron correlations in
oxide materials give rise to a rich variety of electronic phases,
which are highly susceptible to small changes of control
parameters. Although this situation is ideal for applications, the
potential of all-oxide electronic devices had been questioned
until very recently. Indeed, it was thought that the chemical
complexity of most oxides would yield devices much inferior to
those based on conventional semiconductors \cite{Oxide:Takagi}. A
paradigm-shift came when Ohtomo and Hwang demonstrated
unprecedented control of the complex LaAlO$_3$/SrTiO$_3$ interface
\cite{LAO:Hwang}, leading to the formation of a high-mobility
electron gas \cite{Thiel}. The carrier density and sheet
conductivity of the interface 2DEG react sensitively to gate
fields and it was successfully patterned on the nano-scale
\cite{switch:Cen}, which is central to the development of oxide
electronics~\cite{Mannhart:interfaces}. However, a full
understanding of the origin of this 2DEG remains elusive. The two
main contenders are oxygen vacancies at the interface
\cite{Ovacancy:Beasley,Ovacancy:Kalabukhov}, and an electronic
reconstruction to avoid a polar-catastrophe
\cite{LAO:Hwang,polar_cat:Hwang}. Distinguishing between these
mechanisms is an essential step in the development of a new
generation of all-oxide devices.

In this letter, we show that a similar 2DEG can be created at the
bare, unreconstructed SrTiO$_3$ surface. Furthermore, we
demonstrate control of its carrier density through exposure to
intense UV radiation. Fig. 1 shows ARPES data from the cleaved
$(001)$ surface of 0.1\% La-doped SrTiO$_3$(001) following
exposure to UV synchrotron light. At least two electron-like band
dispersions can be observed from the ARPES data (Fig. 1a,b). The
shallower band, with Fermi wave number $k_F$ = 0.12~\AA$^{-1}$,
and deeper band, with $k_F$ = 0.175~\AA$^{-1}$, have their band
bottoms situated $\sim$110 and 216 meV below the Fermi level,
respectively.We have investigated the dimensionality of the
induced electronic system by varying the photon energy, and thus
probing the band dispersion along $k_z$ (surface normal). Fig. S2
(supplementary information) shows that the states have negligible
dispersion along $k_z$. This is the defining property of a
two-dimensional electronic state whose wavefunction is confined
along the $z$-direction to a layer of comparable thickness to the
Fermi wavelength. Also, these bands cannot be associated with the
bulk electronic structure: the photon energy used to record the
data shown in Fig. 1 corresponds to $k_z$ of approximately
3.2$\pi/a$ \cite{STO:Non}, close to the Brillouin zone boundary
where no bulk bands exist in the vicinity of the Fermi energy,
even for samples with more than an order of magnitude higher bulk
carrier density than those measured here. Consequently, we
attribute these states to a surface 2DEG.

In order to obtain the surface charge density, we extract the
Luttinger area from the Fermi surface map shown in Fig. 1(c). Two
concentric Fermi surface sheets are observed, corresponding to the
two dispersions in Fig. 1(a). The intensity variation across the
measured Fermi surface is due to pronounced matrix element
effects. The charge density $n_{2D}$ from each concentric sheet
can then be estimated by $n_{2D} =k_F^2/2\pi$, allowing the total
surface charge density to be determined as $7.1\pm 2 \times
10^{13}$ cm$^{-2}$. This value falls within the range of the 2DEG
densities observed at LaAlO$_3$/SrTiO$_3$ interfaces
\cite{scinterface:Reyren, Ovacancy:Beasley, Ovacancy:Kalabukhov}.
The effective masses extracted from parabolic fits to the two
measured dispersions (Fig. 1a) yield surprisingly low values of
0.5 - 0.6~m$_e$, substantially lower than the bulk band masses.
This will contribute to the high electron mobility
\cite{LAO:Hwang}.

Furthermore, as shown in Fig. 2, we find that the 2DEG density is
not fixed to one particular value, but can be varied by exposure
of the cleaved surface to different irradiation doses of UV light.
We found no evidence of any 2DEG states from initial ARPES
measurements of the freshly cleaved surface, suggesting that the
2DEG only forms after exposure to UV light. Following exposure to
55~eV UV light for an irradiation dose of $\sim$32~J/cm$^2$ (Fig.
2a), we observe a single shallow band with $k_F \sim 0.1$~
\AA$^{-1}$ and a band bottom $\sim60$~meV below E$_F$. Upon
increasing the irradiation dose, this band moves downwards, to
higher binding energies, with a second band becoming visible
between the first band and the Fermi level. The $k_F$ positions
and occupied bandwidth of these two bands continue to increase
slowly with further increase in irradiation dose. The
corresponding surface charge densities extracted from the measured
bands are plotted in Fig. 2f, revealing a monotonic increase in
2DEG density with increasing irradiation dose. Therefore, the
method utilised here provides a controllable means with which to
modify the 2DEG density.

We note that Fig. 2g and 2h, which are measured immediately after
Fig. 2b and 2e, respectively, but with lower intensity of the
probing beam (I = 0.06 W/cm$^2$), show identical band dispersions
within experimental accuracy. Similar results have been obtained
for spectra taken up to an hour after irradiation with an intense
UV beam (not shown). This demonstrates that the 2DEG reported here
is a ground state property of the UV-irradiated SrTiO$_3$ surface.
Hence, its origin must be fundamentally different from
photocarrier doping effects, which have characteristic lifetimes
of the excited states $< 1$~ms at low temperature
\cite{photodoped:Hwang}.

It is clear, therefore, that the irradiation by UV light mediates
a change in the surface of the SrTiO$_3$, which consueqntly
induces the 2DEG. Clear ($1 \times 1$) low-energy electron
diffraction (LEED) patterns observed both before and after the UV
exposure indicate that the surface does not reconstruct during
this process (see Supplementary Fig. S1). Therefore, creation of a
surface 2DEG brought about by a change in the intrinsic surface
state distribution due to surface reconstruction can be ruled out.
This indicates that extrinsic states, such as donor-like defects
or adsorbates, induce the 2DEG. In particular, oxygen vacancies
localized at the surface would be expected to lead to a surface
electron accumulation, with charge neutrality requiring the
creation of an electron 2DEG to screen the positive surface charge
of such ionized defect centres. Experimental studies on
LaAlO$_3$/SrTiO$_3$ interfaces \cite{Ovacancy:Beasley,
Ovacancy:Kalabukhov} have shown that oxygen vacancies can be
created during the sample preparation process under low oxygen
pressures. Here, we suggest that exposure to intense UV light in
ultra-high vacuum causes oxygen desorption from the surface. Such
a photon-induced chemical change was previously observed in
photoluminescence spectra of SrTiO$_3$ following irradiation with
325nm laser light \cite{PL}, where a sub-band-gap luninescence
peak, growing in magnitude with increasing irradiation, and stable
for some time following the irradiation, was assigned to
photo-induced oxygen vacancies. This is consistent with an
increased in-gap defect state \cite{LaSTO:Aiura} that we observe
at $\sim1.3$eV below the Fermi level in angle-integrated
photoemission spectra following the UV exposure (see Supplementary
Information).

We cannot exclude that this in-gap state could also arise from
adsorbed impurities such as hydrogen, which could themselves
provide the required donor-type surface states, as discussed in
the supplementary information. However, irrespective of the exact
microscopic identification of the defects causing the charge
accumulation, the results presented above demonstrate that it is
not necessary to have an interface with the polar surface of
another material in order to obtain a 2DEG at the surface of
SrTiO$_3$. Indeed, we find that extrinsic mechanisms are
sufficient to induce a ground-state 2DEG of the same density.
Apart from obvious advantages for their spectroscopic
investigation, the methodology employed here to create such a 2DEG
offers potential for the realization of novel schemes in oxide
electronics. While existing approaches have the ability to pattern
the spatial extent of the LaAlO$_3$/SrTiO$_3$ interface 2DEG
\cite{switch:Cen,Efield:Caviglia}, our work opens the way to
spatial control of its ground state density by employing focussed
light. Furthermore, UV interference patterns could be employed to
allow much faster parallel nano-scale patterning of the 2DEG. This
should not be specific to the surface of SrTiO$_3$, and could be
employed for creation of surface 2DEGs across a range of oxide
materials. The scheme should also be useful to write surface
charge on LaAlO$_3$ by desorbing oxygen. For a thin layer of
LaAlO$_3$ grown on SrTiO$_3$, charge localized at the surface of
the LaAlO$_3$ is thought to provide the mechanism for writing of
metallic lines at the LaAlO$_3$/SrTiO$_3$ interface using
conducting atomic-force microscopy~\cite{AFM_origin}. Writing of
surface charge on the LaAlO$_3$ layer using UV interference
patterns could therefore be an attractive route towards efficient
processing of a high-mobility modulation-doped 2DEG, patterned at
the nano-scale.

To further characterize the 2DEG created here, we adapt a model
originally developed for conventional semiconductors
\cite{PSCal:Phil}. The charge resulting from surface- (or indeed
interface-) localised oxygen vacancies induces a spatial
redistribution of bulk carriers in the vicinity of the
surface/interface, correlated with a bending of the electronic
bands relative to the Fermi level. If the potential well created
by this band bending is sufficiently deep, it causes the
conduction band states to become quantized into two-dimensional
(2D) subbands. We have performed coupled Poisson-Schr{\"o}dinger
calculations \cite{PSCal:Phil} for such a band-bending scenario.
Incorporating an electric-field dependence of the susceptibility
\cite{ESus:Barthelemy} within our model, we find that the downward
band bending in SrTiO$_3$ (Fig. 3b) is indeed very rapid. This
leads to a narrow 2DEG (Fig. 3c) in a 3D crystal due only to the
strength of the internal electric field. Its lowest subband is
localized within $\approx4$ unit cells from the surface and is
followed by a series of higher subbands with wavefunctions that
progressively extend deeper into the bulk  (see Supplementary
Fig.~S4). As shown in Fig. 3a, these subbands effectively
reproduce the two main dispersions observed in the ARPES data,
confirming that the states observed here can broadly be described
as the quantum well states of a surface 2DEG resulting from a
downward band bending. The additional weakly bound states found in
the calculations at lower binding energies are not resolved
experimentally. This is possibly due to low matrix elements for
shallow quantum well states, which generally have small amplitudes
of the wave function in the near-surface region probed by ARPES,
as shown in Supplementary Fig.~S4. Further, since the wave
functions of these shallow states expand over a much larger depth
than the more localized deeper states, the shallow states might
start dispersing in $k_z$, causing them to become smeared out in
the ARPES measuerements, possibly beyond our detection limit.

Our model does not take intra-unit cell potential variations into
account, thought to be important for the relatively local Ti 3d
states. Nonetheless, it is in good qualitative agreement with
density-functional theory calculations for the LaAlO$_3$/SrTiO$_3$
interface by Popovi{\'c}~{\it et al.}~\cite{LDA_Popovic}, which
report a similar reconstruction of the electronic structure in the
2DEG into a ladder of subbands. The lowest of these was identified
as an in-plane Ti~$d_{xy}$ state largely localized on the
interfacial TiO$_2$ layer, whereas the charge density from a
series of higher-lying $d_{xy}$ states appeared spread out over
several layers in the calculation. Additionally these authors
found $d_{yz}$ and $d_{xz}$ derived states with strongly
elliptical Fermi surfaces and high effective masses for transport
along $x$ and $y$, respectively. In the present case, we observe
only two concentric isotropic Fermi surfaces of light carriers,
which we attribute to the lowest members of a ladder of $d_{xy}$
states. However, we cannot rule out the additional presence of
heavy d$_{xz/yz}$ bands in the surface 2DEG, since their intensity
would be suppressed relative to $d_{xy}$ states in the
experimental geometry and polarization employed here. Also, the
Fermi level in our experiment lies closer to the lowest subband
minimum than in the calculations of Popovi{\'c} {\it et
al.}~\cite{LDA_Popovic} for the LaAlO$_3$/SrTiO$_3$ interface, and
so the $d_{xz/yz}$ states, if present, should be quite shallow and
hence extend deep in the bulk, which would reduce their intensity
in ARPES.

Although most conventional semiconductors exhibit a depletion of
charge carriers at the surface, a small number have themselves
been observed to support a surface 2DEG, concomitant with a
downward band bending  \cite{In2O3:Phil}. Comparing the 2DEG
states from one such example, InAs \cite{Phil:SBs} shown in Fig.
4b, with those of SrTiO$_3$ observed here (Fig. 4a), reveals a
qualitative similarity between the two materials. This further
confirms the validity of the above model. However, some important
differences are also apparent. In particular, there is pronounced
spectral weight in SrTiO$_3$ at binding energies much higher than
the band bottom. This effect is absent for the semiconductor case,
but can be seen in ARPES measurements of other strongly-correlated
compounds such as the cuprate high-temperature superconductor
Bi$_2$Sr$_2$CuO$_{6}$ (Fig. 4c) \cite{HEA:Non}. The non-vanishing
spectral weight in SrTiO$_3$ implies a finite electron self-energy
at high binding energies, giving direct evidence of enhanced
many-body interactions inherent to the 2DEG states of SrTiO$_3$.
Hence, the 2DEG here is best described as an electron liquid
rather than an electron gas \cite{Mannhart:2DEL}. Besides this
spectral weight below the band bottom, we also note that the
SrTiO$_3$ data show weak dispersion anomalies (kinks) at a binding
energy around 20-30 meV and, less clearly, around 70-80 meV below
the Fermi level, which we assign to electron-phonon interactions
with a weaker coupling strength than observed in the bulk
\cite{STO:Non}.

In systems where electronic correlations play an important role,
the quasiparticles are normally found to be heavy. Intriguingly,
however, the strong electron correlations we observe here do not
lead to a substantial mass enhancement within the 2DEG. In fact,
the effective mass of 0.5-0.6m$_e$ extracted from the data is
lower than the lightest bulk band mass of $\sim$m$_e$ estimated
from ref. \cite{STO:Non}. Almost certainly, this will apply to the
LaAlO$_3$/SrTiO$_3$ interface 2DEG too, which helps to explain the
high electron mobilities achieved in this system. While a direct
spectroscopic measurement of the electronic band dispersion within
the 2DEG, as performed here, has not yet been achieved in the
interface systems, this conclusion is supported by very recent
measurements of the penetration field in front-gated
LaAlO$_3$/SrTiO$_3$ heterostructures, where a 2DEG band mass
significantly below any of  the bulk masses was
inferred~\cite{Ashoori:lightmass}. We speculate that this unusual
behaviour might be due to an interaction-induced shrinkage of the
fundamental band gap approaching the surface of SrTiO$_3$
\cite{Phil:SBs}. This would effectively increase the depth of the
potential well and hence result in steeper quantized bands/lighter
band masses. If this picture is true, the surface of SrTiO$_3$, or
indeed its interface with LaAlO$_3$, are rare examples where
many-body interactions have the counter-intuitive effect of
increasing the mobility. A full understanding of this will require
further theoretical and experimental studies. Spectroscopic
investigations of surface 2DEGs, of the form reported here, will
likely prove essential to elucidate the fundamental electronic
structure and underlying role of many-body interactions in oxide
2DEGs, and so will play a major role in the development of
all-oxide electronics.

%\newpage
%\References

\begin{acknowledgments}
We would like to thank H.Y. Hwang, H. Takagi, M.R. Beasley, J. L.
M. van Mechelen, D. van der Marel, P. Reunchan, and S.
Limpijumnong for helpful discussions. W.M. would like to thank H.
Nakajima and Y. Rattanachai for help with the resistivity
measurement. The work at ALS and Stanford Institute for Materials
and Energy Sciences are supported by DOE's Office of Basic Energy
Sciences under Contracts No. DE-AC02-76SF00515 and
DE-AC03-76SF00098. The work at St Andrews is supported by the
UK-EPSRC (EP/F006640/1) and the ERC (207901). W.M. acknowledges
The Thailand Research Fund, Office of the Higher Education
Commission and Suranaree University of Technology for financial
support.
\end{acknowledgments}

\subsection{Author contributions} ARPES measurements were performed by
W.M., P.D.C.K., R.-H.H., F.B., and A.T.  W.M. and P.D.C.K.
analysed the ARPES data. W.M., P.D.C.K. and F.B. wrote the paper
with suggestions and comments by R.-H.H., S.-K.M. and Z.-X.S.
Calculations of quantized 2DEG states were done by P.D.C.K.
S.-K.M. and M.H. maintained the ARPES endstation. Resistivity
measurements were performed by W.M. and P.S. Z.-X.S. and F.B. are
responsible for project direction, planning and infrastructure.

%%%%%%%%% figures in the main body %%%%%%%%%
\newpage

\begin{figure} [t]
\includegraphics [width=4.5in, clip]{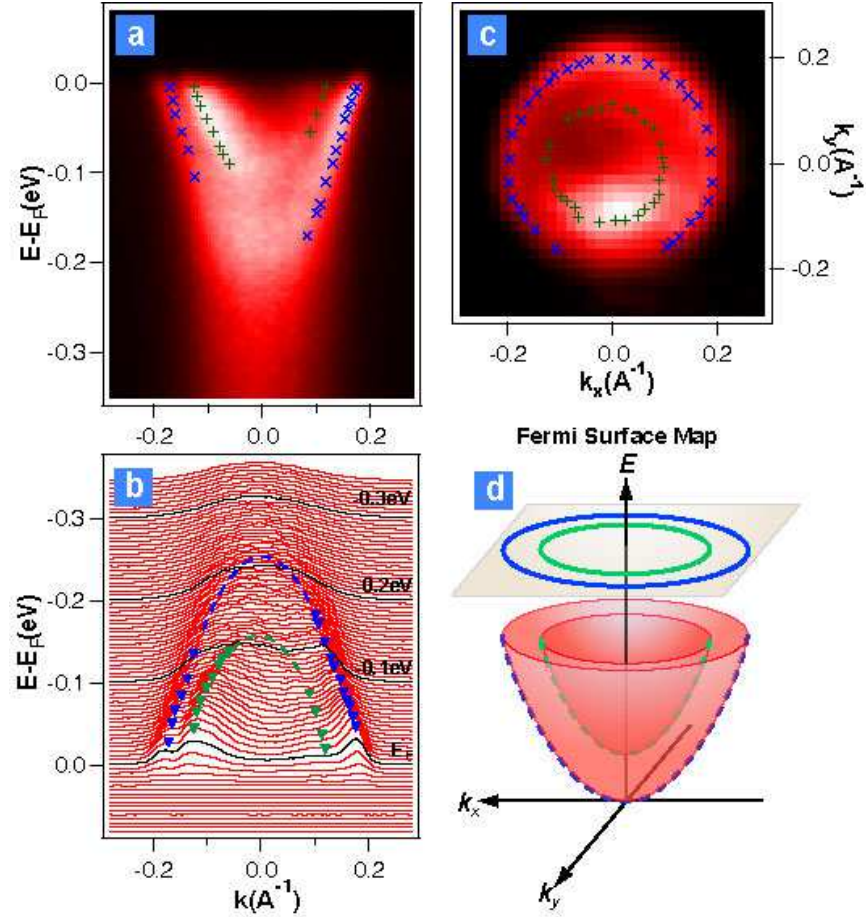}
\caption{\label{fig1} Observation of a surface 2DEG on SrTiO$_3$
after exposure of the cleaved (100) surface to synchrotron (UV)
light. (a) ARPES data of La$_x$Sr$_{1-x}$TiO$_3$ (x=0.001) at T =
20K, with corresponding momentum distribution curves in (b). The
sample has been irradiated with $\approx$480~J/cm$^2$ UV light of
55~eV with an intensity of $\sim$ 0.34W/cm$^2$. The ARPES data are
taken in the second Brillouin zone using the same photon energy.
The dashed lines in (b) are parabolic fits to the data points
(symbols) extracted from the ARPES data; the green and blue curves
have effective masses of $\sim$0.6 and 0.5 m$_e$, respectively.
(c) Fermi surface map, taken on a different sample following the
same preparation. Two concentric circular Fermi surface sheets
(symbols) are visible. (d) shows the schematic Fermi surface and
band dispersions obtained from the measured electronic structure.}
\end{figure}

\begin{figure*}
\includegraphics [width=6in, clip]{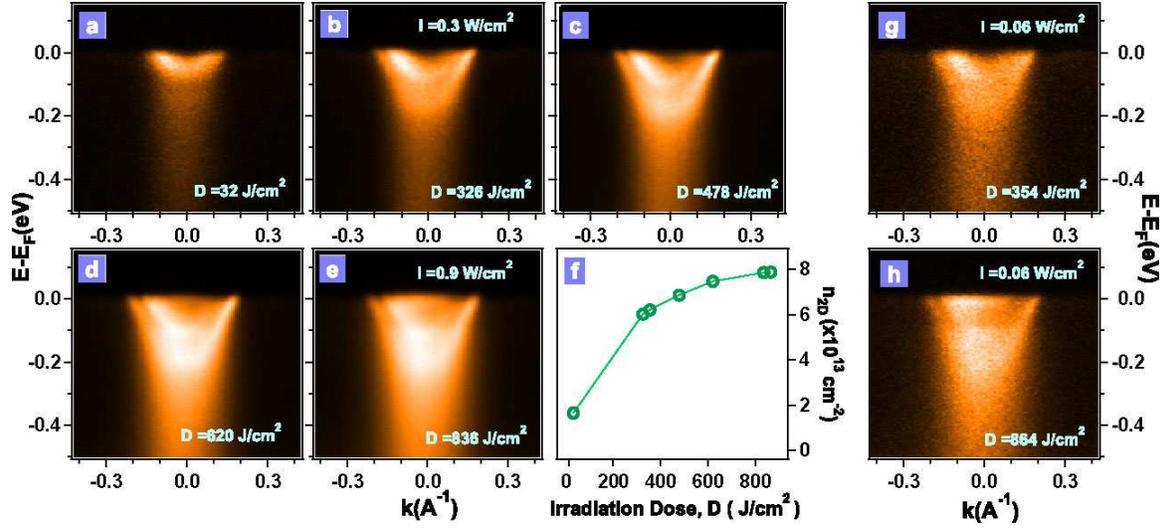}
\caption{\label{fig2} Variation of 2DEG charge density with
exposure to different UV irradiation doses. (a)-(e), (g) and (h)
show ARPES data for different irradiation doses indicated in the
figure. The corresponding 2DEG charge densities as a function of
irradiation dose are shown in (f). (g) and (h) show ARPES data
measured immediately after (b) and (e), respectively, but with
lower intensity of the probing photon beam.}
\end{figure*}

\begin{figure*}
\includegraphics [width=6in, clip]{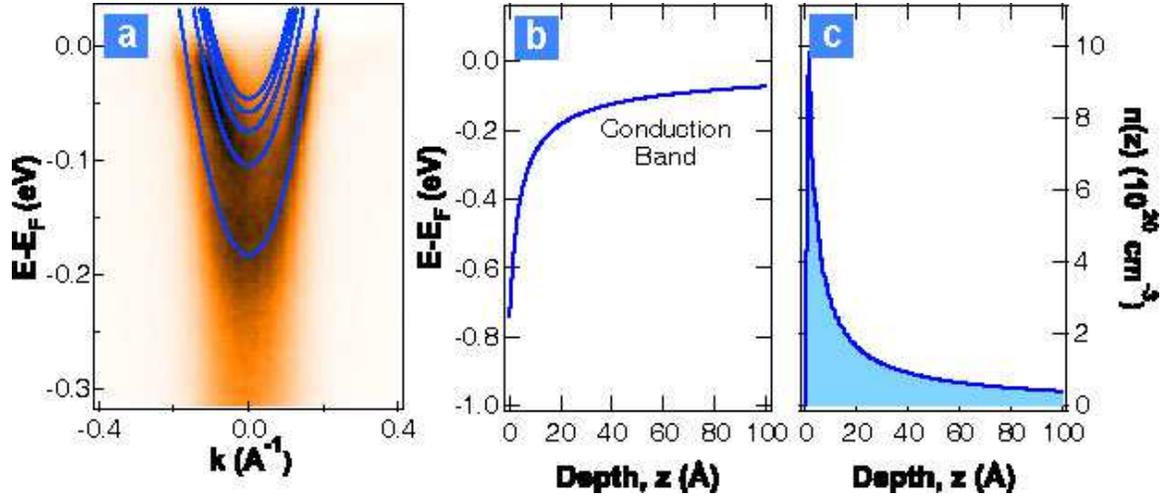}
\caption{\label{fig3:qp} Calculations of quantized 2DEG states
within a band-bending model \cite{PSCal:Phil}. (a),(b) The
calculations yield quantized 2DEG states (solid lines in (a))
inside the potential well caused by the downward bending of the
conduction band minimum (solid line in (b)) relative to the Fermi
level, when approaching the surface of the material. The
corresponding three-dimensional charge density variation as a
function of depth is shown in (c). We note that in the
calculation, there are three additional shallow states which are
not clearly observed in the data. It is possible that these states
exist in the data but are suppressed due to a combination of
matrix element effects and broadening due to finite $k_z$
dispersion of the shallow states. These considerations are
supported by the observations from InAs (Fig. 4b), as well as
other semiconductors \cite{InN:Colakerol,Phil:SBs}, where shallow
states of a surface quantum well are more smeared out than the
deeper ones.}
\end{figure*}

\begin{figure} [t]
\includegraphics [width=5.5in, clip]{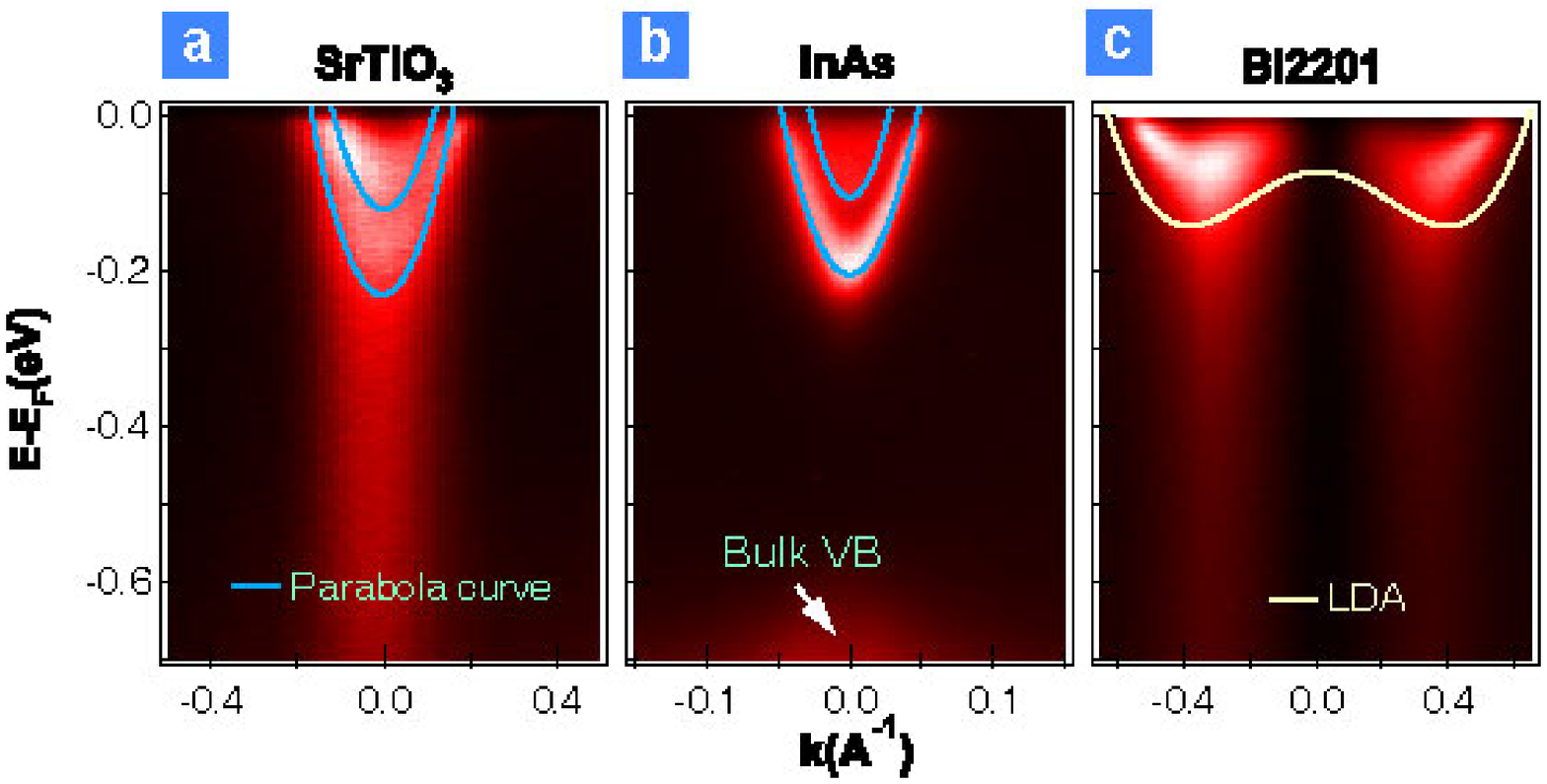}
\caption{\label{fig3:qp} Comparison of ARPES data from SrTiO$_3$,
InAs and Bi$_2$Sr$_2$CuO$_6$ (Bi2201) samples. (a), (b) show 2DEG
states at the surface of SrTiO$_3$ and InAs, respectively. (c)
shows ARPES data from the single layer cuprate Bi2201 along ($\pi,
0$) to (0, $\pi$). Lines in (a) and (b) are parabolic dispersion
relations, to guide the eye, and the result of an LDA
band-structure calculation in (c).}
\end{figure}


\begin{thebibliography}{}
%%%%%%%%% refs in the main body %%%%%%%%%

\bibitem{HighTc:Bednorz}
Bednorz, J.G. \& Muller, K.A. Perovskite-type oxides: The new
approach to high-Tc superconductivity. \emph{Rev. Mod. Phys.}
\textbf{60}, 585-600 (1988).

\bibitem{LMO:Helmolt}
Von Helmolt, R. \emph{et al.} Giant negative magnetoresistance in
perovskitelike La$_{2/3}$Ba$_{1/3}$MnO$_x$ ferromagnetic films.
\emph{Phys. Rev. Lett.} \textbf{71}, 2331-2333 (1993).

\bibitem{multiferroic:Kimura}
Kimura, T. \emph{et al.} Magnetic control of ferroelectric
polarization.  \emph{Nature} \textbf{426}, 55-58 (2003).

\bibitem{LAO:Hwang}
Ohtomo, A. \& Hwang, H.Y. A high-mobility electron gas at the
LaAlO$_3$/SrTiO$_3$ heterointerface. \emph{Nature} \textbf{427},
423-426 (2004).

\bibitem{Oxide:Takagi}
Takagi, H. \& Hwang, H.Y. An Emergent Change of Phase for
Electronics. \emph{Science} \textbf{327}, 1601-1602 (2010).

\bibitem{Thiel}
Thiel, S. \emph{et al.}, Tunable Quasi-Two-Dimensional Electron
Gases in Oxide Heterostructures. \emph{Science} \textbf{313},
1942-1945 (2006).

\bibitem{scinterface:Reyren}
Reyren, N. \emph{et al.}, Superconducting Interfaces Between
Insulating Oxides. \emph{Science} \textbf{317}, 1196-1199 (2007).

\bibitem{magnetic:Brinkman}
Brinkman, A. \emph{et al.} Magnetic effects at the interface
between non-magnetic oxides. \emph{Nature Mater.} \textbf{6},
493-496 (2007).

\bibitem{switch:Cen}
Cen, C. \emph{et al.} Nanoscale control of an interfacial
metal--insulator transition at room temperature. \emph{Nature
Mater.} \textbf{7}, 298-302 (2008).

\bibitem{Mannhart:interfaces}
Mannhart, J. \& Scholm, D. G. Oxide Interfaces --- An Opportunity
for Electronics. \emph{Science} \textbf{327}, 1607-1611 (2010).

\bibitem{Ovacancy:Beasley}
Siemons, W. \emph{et al.} Origin of Charge Density at LaAlO$_3$ on
SrTiO$_3$ Heterointerfaces: Possibility of Intrinsic Doping.
\emph{Phys. Rev. Lett.} \textbf{98}, 196802 (2007).

\bibitem{Ovacancy:Kalabukhov}
Kalabukhov, A. \emph{et al.} Effect of oxygen vacancies in the
SrTiO3 substrate on the electrical properties of the
LaAlO$_3$/SrTiO$_3$ interface. \emph{Phys. Rev. B} \textbf{75},
121404(R) (2007).

\bibitem{polar_cat:Hwang}
Nakagawa, N., Hwang, H. Y. \& Muller, D. A. Why some interfaces
cannot be sharp. \emph{Nature Mater.} \textbf{5}, 204-209 (2006).

\bibitem{STO:Non}
Meevasana, W. \emph{et al.} Strong energy-momentum dispersion of
phonon-dressed carriers in the lightly doped band insulator
SrTiO$_3$. \emph{New. J. Phys.} \textbf{12}, 023004 (2010).

\bibitem{photodoped:Hwang}
Kozuka, Y. \emph{et al.} Optically tuned dimensionality crossover
in photocarrier-doped SrTiO$_3$: Onset of weak localization.
\emph{Phys. Rev. B} \textbf{76}, 085129 (2007).

\bibitem{PL}
Mochizuki, S.\emph{et al.} Photoluminescence and reversible
photo-induced spectral change of SrTiO$_3$. \emph{J.
Phys.:Condens. Matter} \textbf{17}, 923-948 (2005).

\bibitem{LaSTO:Aiura}
Aiura, Y. \emph{et al.} Photoemission study of the metallic state
of lightly electron-doped SrTiO$_3$. \emph{Surf. Sci.}
\textbf{515}, 61-74 (2002), \emph{and references therein}.

\bibitem{Efield:Caviglia}
Caviglia, A. D. \emph{et al.} Electric field control of the
LaAlO$_3$/SrTiO$_3$ interface ground state. \emph{Nature}
\textbf{456}, 624-627 (2008).

\bibitem{AFM_origin}
Xie, Y. \emph{et al.} Charge Writing at the LaAlO$_3$/SrTiO$_3$
Surface. \emph{Nano Lett.} \textbf{10}, 2588-2591 (2010).

\bibitem{PSCal:Phil}
King, P. D. C., Veal, T. D. \& McConville, C. F. Non-parabolic
coupled Poisson-Schrodinger solutions for quantized electron
accumulation layers: Band bending, charge profile, and subbands at
InN surfaces. \emph{Phys. Rev. B} \textbf{77}, 125305 (2008).

\bibitem{ESus:Barthelemy}
Copie, O. \emph{et al.} Towards Two-Dimensional Metallic Behavior
at LaAlO$_3$/SrTiO$_3$ Interfaces. \emph{Phys. Rev. Lett.}
\textbf{102}, 216804 (2009).

\bibitem{LDA_Popovic}
Popovi{\'c}, Z. S., Satpathy, S. \& Martin, R. M. Origin of the
Two-Dimensional Electron Gas Carrier Density at the LaAlO$_3$ or
SrTiO$_3$ interface. \emph{Phys. Rev. Lett.} \textbf{101}, 256801
(2008).

\bibitem{In2O3:Phil}
King, P.D.C. {\emph et~al.}, Surface Electron Accumulation and the
Charge Neutrality Level in In$_2$O$_3$. {\emph{Phys. Rev. Lett.}}
{\textbf{101}}, 116808 (2008), \emph{and references therein}.

\bibitem{Phil:SBs}
King, P. D. C. \emph{et al.} Surface band gap narrowing in
quantized electron accumulation layers. \emph{Phys. Rev. Lett.}
{\textbf{104}}, 256803 (2010).

\bibitem{InN:Colakerol}
Colakerol, L. \emph{et al.} Quantized Electron Accumulation States
in Indium Nitride Studied by Angle-Resolved Photoemission
Spectroscopy. \emph{Phys. Rev. Lett.} \textbf{97}, 237601 (2006).

\bibitem{HEA:Non}
Meevasana, W. \emph{et al.} Hierarchy of multiple many-body
interaction scales in high-temperature superconductors.
\emph{Phys. Rev. B} \textbf{75}, 174506 (2007).

\bibitem{Mannhart:2DEL}
Breitschaft, M. \emph{et al.} Two-dimensional electron liquid
state at LaAlO$_3$-SrTiO$_3$ interfaces. \emph{Phys. Rev. B}
\textbf{81}, 153414 (2010).

\bibitem{Ashoori:lightmass}
Li, L. \emph{et al.} Large capacitance enhancement and negative
compressibility of two-dimensional electronic systems at
LaAlO$_3$/SrTiO$_3$ interfaces. Preprint at
$<$http://arxiv.org/pdf/1006.2847$>$ (2010).

\end{thebibliography}
\end{document}